\documentclass[twocolumn]{aastex631}

\shorttitle{Hidden AGN in bipolar blobs galaxy}
\shortauthors{Yao et al.}

\begin{document}

\title{Bipolar blobs as evidence of hidden AGN activities in the low-mass galaxies}

\correspondingauthor{Enci Wang, Zhicheng He, and Xu Kong}
\email{ecwang16@ustc.edu.cn, zcho@ustc.edu.cn, xkong@ustc.edu.cn}

\author[0000-0002-6873-8779]{Yao Yao}
\affiliation{Deep Space Exploration Laboratory / Department of Astronomy, University of Science and Technology of China, Hefei, Anhui, 230026, China}
\affiliation{School of Astronomy and Space Sciences, University of Science and Technology of China, Hefei, Anhui, 230026, China}

\author{Enci Wang}
\affiliation{Deep Space Exploration Laboratory / Department of Astronomy, University of Science and Technology of China, Hefei, Anhui, 230026, China}
\affiliation{School of Astronomy and Space Sciences, University of Science and Technology of China, Hefei, Anhui, 230026, China}

\author{Zhicheng He}
\affiliation{Deep Space Exploration Laboratory / Department of Astronomy, University of Science and Technology of China, Hefei, Anhui, 230026, China}
\affiliation{School of Astronomy and Space Sciences, University of Science and Technology of China, Hefei, Anhui, 230026, China}

\author{Zheyu Lin}
\affiliation{Deep Space Exploration Laboratory / Department of Astronomy, University of Science and Technology of China, Hefei, Anhui, 230026, China}
\affiliation{School of Astronomy and Space Sciences, University of Science and Technology of China, Hefei, Anhui, 230026, China}

\author{Yu Rong}
\affiliation{Deep Space Exploration Laboratory / Department of Astronomy, University of Science and Technology of China, Hefei, Anhui, 230026, China}
\affiliation{School of Astronomy and Space Sciences, University of Science and Technology of China, Hefei, Anhui, 230026, China}

\author{Hong-Xin Zhang}
\affiliation{Deep Space Exploration Laboratory / Department of Astronomy, University of Science and Technology of China, Hefei, Anhui, 230026, China}
\affiliation{School of Astronomy and Space Sciences, University of Science and Technology of China, Hefei, Anhui, 230026, China}

\author{Xu Kong}
\affiliation{Deep Space Exploration Laboratory / Department of Astronomy, University of Science and Technology of China, Hefei, Anhui, 230026, China}
\affiliation{School of Astronomy and Space Sciences, University of Science and Technology of China, Hefei, Anhui, 230026, China}



\begin{abstract}

We report the evidence of a hidden black hole (BH) in a low-mass galaxy, MaNGA 9885-9102, and provide a new method to identify active BH in low mass galaxies.  
This galaxy is originally selected from the MaNGA survey with distinctive bipolar H$\alpha$ blobs at the minor axis.
The bipolar feature can be associated with AGN activity, while the two blobs are classified as the H II regions on the BPT diagram, making the origins confusing.  
The Swift UV continuum shows that the two blobs do not have UV counterparts, suggesting that the source of ionization is out of the blobs. Consistent with this, the detailed photoionization models prefer to AGN rather than star-forming origin with a significance of 5.8$\sigma$. The estimated BH mass is $M_{\rm BH}\sim$7.2$\times 10^5 M_\odot$ from the $M_{\rm BH}-\sigma_*$ relationship. This work introduces a novel method for detecting the light echo of BHs, potentially extending to intermediate mass, in low metallicity environments where the traditional BPT diagram fails. 

\end{abstract}



\section{Introduction} \label{sec:intro}

Ionized gases are widely present in galaxies and are closely related to star formation and the activity of the central supermassive black holes \citep[SMBHs, ][]{bpt,Kennicutt1998,Heckman1990,Kewley2001}. With the emergence of integral field spectroscopy (IFS) surveys, we can obtain a large number of spatially resolved spectra, where we can find some distinctive ionized regions with unusual physical properties. Recently, some off-galaxy emission-line regions without significant optical counterparts (i.e. H$\alpha$ blobs) are detected in local IFS surveys \citep{Lin2017,Bait2019,Pan2020,Ji2021}. These blobs have been rarely observed and studied, and their physical origin is still unclear. Some of them are believed to be possibly related to the Active galactic nuclei \citep[AGNs, ][]{Bait2019,Pan2020}. Finding H$\alpha$ blobs may gives us a new way to investigate AGN activities, especially for weak, obscured, and faded AGNs \citep[e.g. Hanny's Voorwerp, ][]{Lintott2009}. Furthermore, it can also become a potential method for detecting intermediate mass black holes (IMBHs) in dwarf galaxies, which are intriguing in enhancing our comprehension of the historical growth and evolution of all central supermassive black holes  \citep{fryer2001,madau2001,Gueltekin2022}.

When a black hole is obscured or dormant, its presence can be inferred through the ionization region at the galactic scale, known as the BPT (named after ``Baldwin, Phillips \& Telervich") diagnostic map \citep{bpt}. However, this traditional BPT diagram may not work for IMBH potential hosts because of the typical low metallicity of host galaxies, especially for galaxies with stellar masses less than $10^{10} M_\odot$.

Here, we report a low-mass galaxy initially misclassified as an H II region by traditional BPT analysis. However, by recognizing its distinctive biconical ionization region \citep{antonucci1993,he2018,shen2023}, and combining its multi-band properties, we confirm the presence of a black hole at its center, whose mass ($\sim$7.2$\times 10^5M\odot$) is close to the upper limit of IMBH.

This paper is organized as follows. In Sect. \ref{sec:identification}, we introduce the identification of this galaxy, including the X-ray observations. Sect. \ref{sec:result} presents the our data analysis and the main results from both observations and models. Sect. \ref{sec:discuss} discusses the physical origin of the two blobs, with the comparison between the different ionization models. Finally, in Sect. \ref{sec:conclusion}, we conclude the paper with a summary and their significance. Throughout this paper, we adopt a Lambda cold dark matter cosmology with $\Omega_\Lambda$=0.7, $\Omega_m$=0.3, and $H_0$=70 km s$^{-1}$ Mpc$^{-1}$.

\section{The galaxy identification} \label{sec:identification}

\subsection{Morphological selection in MaNGA}

\begin{figure*}[ht]
\centering
\includegraphics[width=1\textwidth]{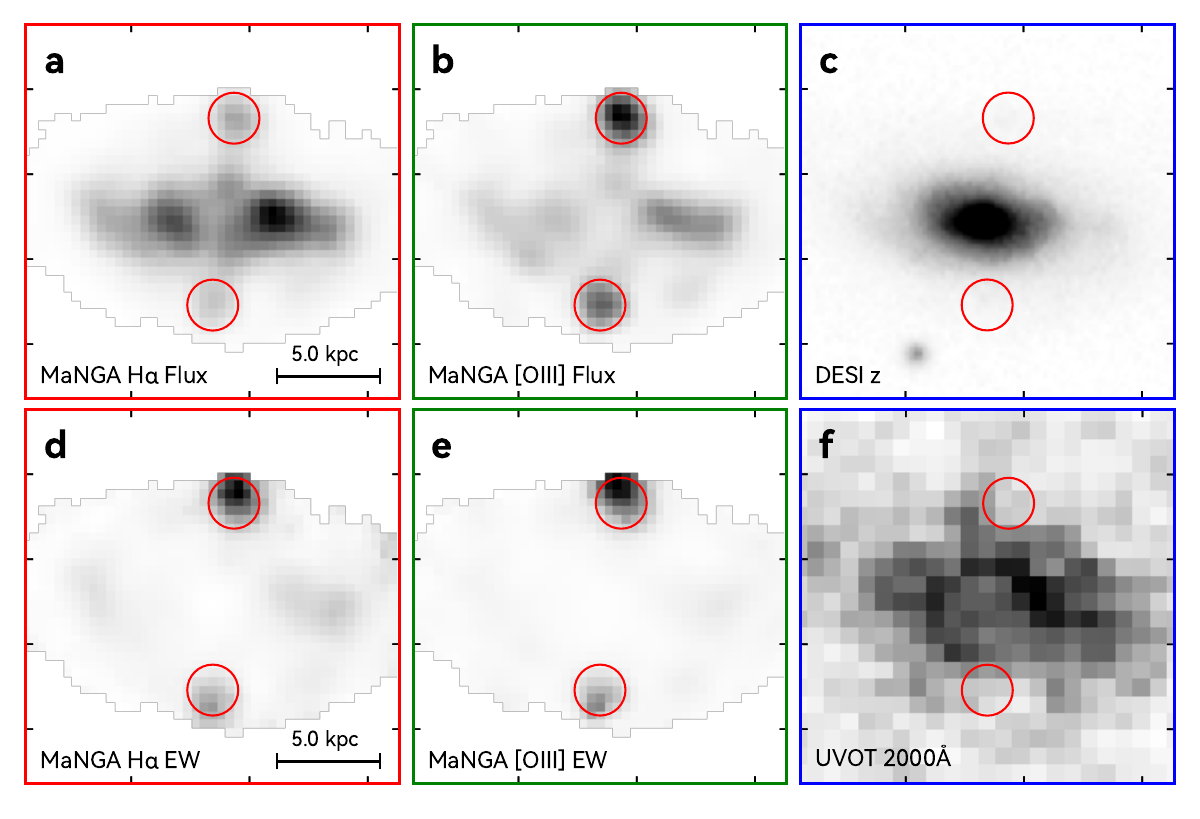}
\caption{Images of MaNGA 9885-9102. The first column (Panel a and d) shows the maps of H$\alpha$ flux and equivalent width. The second column (Panel b and e) shows the maps of [O III]$\lambda$5007. The third column (Panel c and f) shows the images of the DESI z band and UVOT UVW2 band. The red circles indicate the positions of the upper and lower blobs, and their sizes represent the area of the blobs. }\label{fig:main_image}
\end{figure*}

This galaxy is from Mapping Nearby Galaxies at Apache Point Observatory (MaNGA) survey \citep{Bundy2015}, named as MaNGA 9885-9102. It is a disk and star-forming galaxy with stellar mass of 4.4$\times 10^9 M_\odot$ at redshift of 0.0417. It is originally selected from the investigation of non-parametric morphology \citep{Conselice2014} of emission line flux map from the Data Analysis Pipeline (DAP) of MaNGA \citep{dap,dap_emission} galaxies \citep{statmorph, statmorph_csst}. We conduct a manual visual inspection of the galaxies with special features. The most evident feature of this galaxy is the bi-polar H$\alpha$ and [O III]$\lambda$5007 blobs at the minor axis, as shown in Fig. \ref{fig:main_image}a and \ref{fig:main_image}b (also see Fig. \ref{fig:main_image}d and Fig. \ref{fig:main_image}e for equivalent width). Interestingly, by examining the deeper broad-band image of this galaxy from DESI legacy imaging survey, we find no counterparts on the $z$-band image for the two blobs, and only two dots in the $g$-band contributed by the flux of [O III]$\lambda$5007. This indicates that no old stellar populations reside in the two blobs, rejecting the possibility of coincidence of two dwarf galaxies located at the symmetric positions of MaNGA 9885-9102.

\begin{figure*}[ht]
\centering
\includegraphics[width=1\textwidth]{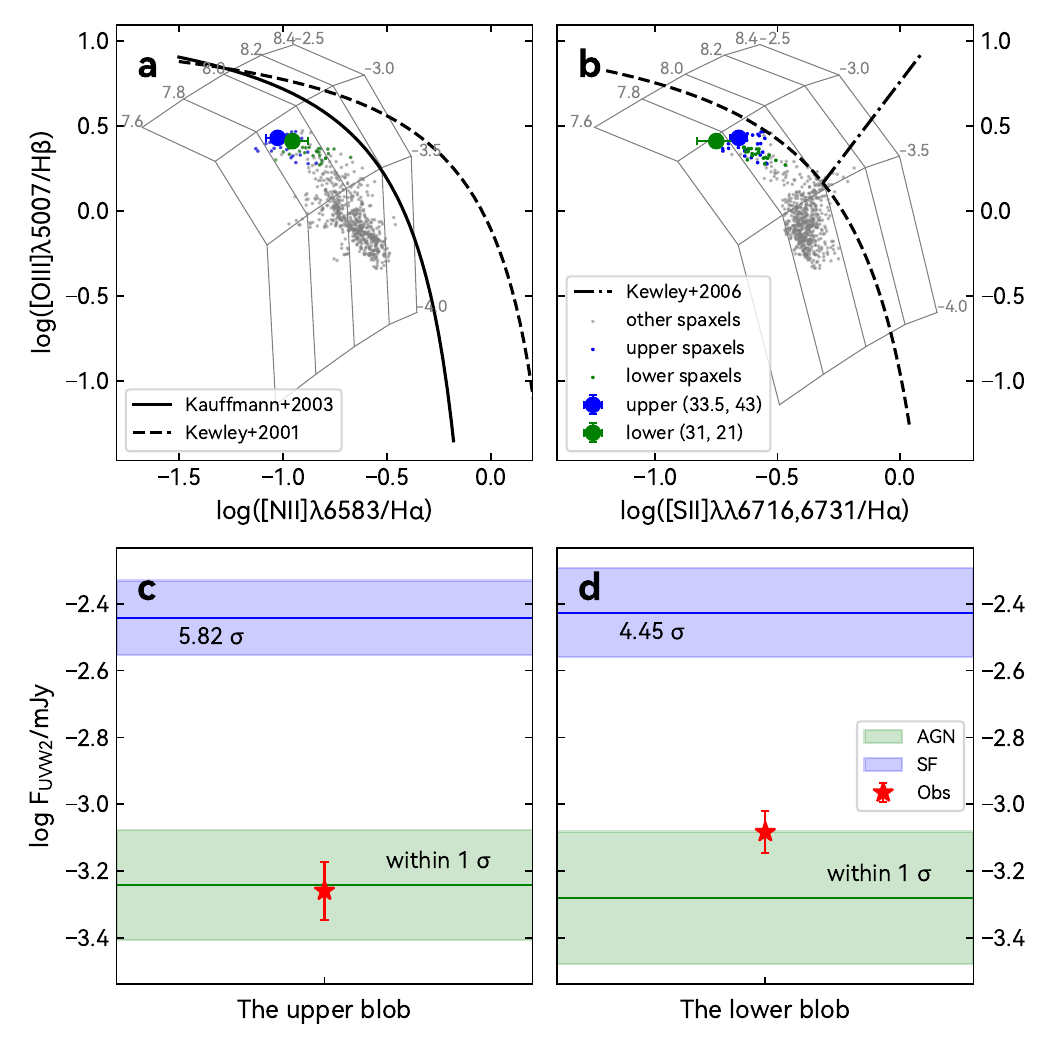}
\caption{Panel a and b: [N II] and [S II] BPT diagrams of MaNGA 9885-9102. The solid and dashed black curves are the empirical maximum starburst line of \cite{Kauffmann2003} and \cite{Kewley2001} respectively. The dash-dotted black line in the [S II] diagram is the boundary between Seyfert and LINER \citep{Kewley2006}. The line ratios of spaxels in the upper and lower blob are shown as blue and green small dots respectively. The big dots with error bars are the line ratios of the total flux of each blob. The gray small dots are the line ratios of other spaxels with the S/N of H$\alpha$ and H$\beta$ both greater than 5. Gray grids are generated by photoionization model with the MF87 \citep{mf87} incident spectrum, assuming log(N/O)=$-$1.0 (see Sect. \ref{subsec:cloudy}). Panel c and d: the observed UV flux of the two blobs compared with the UV flux predicted by the star-forming photoionization model and the diffused AGN photoionization model. The values of $\sigma$ in the areas are the deviations between the model and the observed fluxes.}\label{fig:bpt_uv}
\end{figure*}

Previous studies \citep{Lin2017,Bait2019,Pan2020,Ji2021} have suggests various mechanisms of the H$\alpha$ blobs, including tidal remnants from merger systems, SF region of faint spiral arms, and transient gas blob expelled and illuminated by AGN. However, none of them show clear bipolar signature in its appearance. In our case, the well-defined bipolar feature can be interpreted as the AGN feedback. However, these two blobs are classified as the H II regions on BPT diagram as shown in Fig. \ref{fig:bpt_uv}a and \ref{fig:bpt_uv}b, making us unsure whether the physical origin of the blobs is AGN or star formation.

\begin{deluxetable}{cc}
\tabletypesize{\scriptsize}
\tablewidth{0pt} 
\tablecaption{Basic parameters of MaNGA 9885-9102} \label{tab:info}
\tablehead{\colhead{Parameter} & \colhead{Value}}
\colnumbers
\startdata 
PLATE-IFU & 9885-9102 \\
Alternative name & LEDA 1658319 \\
R.A. & 239.6652 \\
Decl. & 21.8381 \\
Redshift & 0.0417 \\
b/a & 0.47 \\
R$_e$/kpc & 5.59 \\
log M$_*$/M$_\odot$ & 9.64 \\
log SFR/(M$_\odot$ yr$^{-1}$) & -0.19 \\
\enddata
\tablecomments{All basic parameters except for the SFR are from the MaNGA DRPall catalog \citep{drp} and the Simbad database. The b/a, R$_e$, log M$_*$ is from the NSA Sersic fitting. The SFR is from the Pipe3D catalog \citep{pipe3d} and based on the H$\alpha$. }
\end{deluxetable}

The basic properties of this galaxy are listed in Table \ref{tab:info}, including minor-to-major axis ratio ($b/a$), effective radius ($R_{\rm e}$), stellar mass and SFR. All these basic parameters except for the SFR are from the MaNGA DRPall catalog \citep{drp} and the Simbad database. The SFR is from the MaNGA Pipe3D catalog \citep{pipe3d} based on the H$\alpha$.

\begin{figure*}[ht]
\centering
\includegraphics[width=1\textwidth]{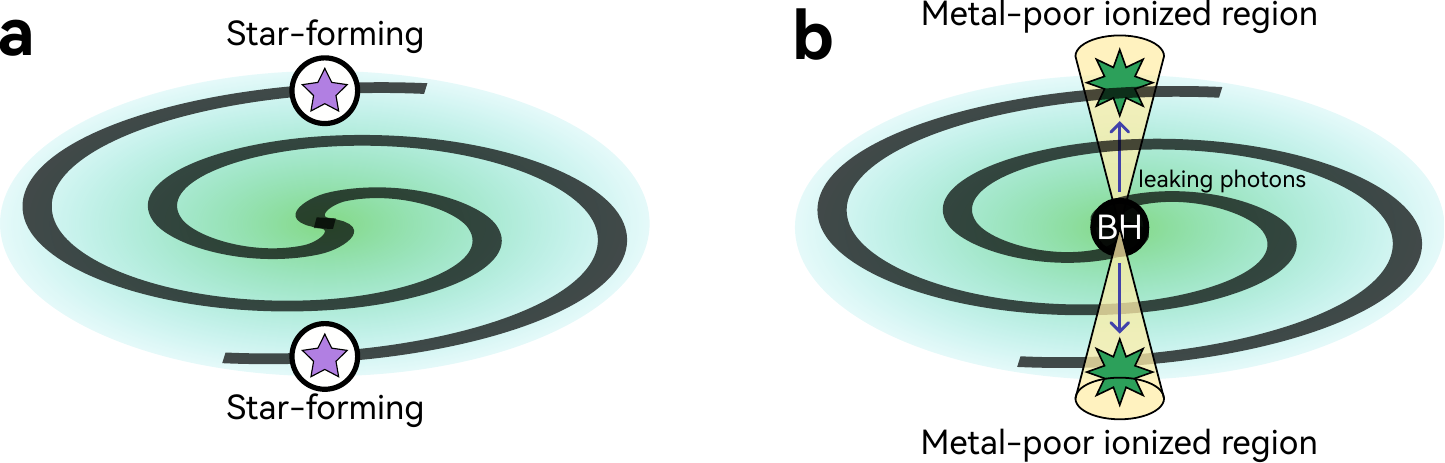}
\caption{The schematic diagram of two type of models for blobs. Panel a: photoionization model from star formation. Panel b: photoionization model from AGN.
}\label{fig:shiyitu}
\end{figure*}

Consequently, We suggest two mechanisms in Fig. \ref{fig:shiyitu} to explain the origin of the blobs. In Fig.\ref{fig:shiyitu}a, the two geometry symmetrical blobs are coincidentally star-forming regions located on the outer faint spiral arms.  Gas is ionized by in-situ young blue O and B stars, producing H II regions and emitting various emission lines in optical band. In Fig. \ref{fig:shiyitu}b, the source of ionization for the blobs is not within the blobs, but the photons emitted by the activity of the central black hole leak out from the central region. The later is similar to the intergalactic cloud Hanny's Voorwerp illuminated by the AGN of IC 2497 \citep{Lintott2009,Jozsa2009,Keel2012,Fabbiano2019}.

\subsection{Swift/XRT and UVOT observations} \label{subsec:swift}

To explore the potential X-ray and UV emission, we submitted a Swift Target-of-Opportunity (ToO) request (ToO ID: 19176, PI: Lin). Two observations with a total exposure time of 3.8 ks were performed, simultaneously on the X-Ray Telescope \citep[XRT, ][]{Burrows2005} and the Ultra-Violet/Optical Telescope \citep[UVOT, ][]{Roming2005}, from August 4th to 9th, 2023. The energy range of XRT is 0.3-10 keV and the band of UVOT is UVW2 (centered at 2000\AA).

For the XRT data, we use \texttt{xrtpipeline} to obtain level 2 files and use \texttt{xrtproducts} to get the level 3 files. The source region is a circle of 20'' around the optical position, while the background region is selected as a source-free annulus with an inner and outer radius of 120'' and 200'', respectively. We use the Bayesian \citep{Kraft1991} method to calculate the 3-$\sigma$ upper limit of the 0.3-10.0 keV photon count rate. After that, we convert the rate into the flux by the online \texttt{WebPIMMS} tool\footnote{\url{https://heasarc.gsfc.nasa.gov/cgi-bin/Tools/w3pimms/w3pimms.pl}}. Since no X-ray photon lies in the source region, a typical AGN spectrum is adopted, which is an absorbed power-law spectrum with an index of $\Gamma=1.75$ \citep{Ricci2017} and a Galactic hydrogen density of $N_{\rm H}=4.48\times10^{20}$ cm$^{-2}$ \citep{HI4PI2016}. The 3-$\sigma$ upper limit of the unabsorbed 0.3-10.0 keV luminosity is $4.3\times10^{41}$ erg s$^{-1}$.

During the main XRT task, UVOT was working in the UVW2 band most of the time, covering 3.4 ks of the exposure time. For the UVW2 data, we first examine each image file and exclude the extensions with bad photometric flags. After that, we stack the images by \texttt{uvotimsum}. We mark the blob regions on the image and find the intensity is much weaker than that of the host galaxy, as seen in Fig.\ref{fig:main_image}f.

Although our main interest lies in the blob regions rather than the host galaxy, we still need to get a conversion factor between the UVW2 luminosity and the background-subtracted and corrected count (noted as $F_{\rm UVW2}$), so that we can get the UVW2 luminosity for the blobs. Therefore, we use \texttt{uvotsource} to perform the photometry for the host galaxy and get these two parameters. The source region is selected as a circle with a radius of 10'' around the galaxy, while the background region is selected as a source-free circle with a radius of 40''. We correct the Galactic extinction by applying the O'Donnell's law \citep{Odonnell1994} and an extinction of $E(B-V) = 0.072$ mag \citep{Schlegel1998}, which is the same correction process as the MaNGA DAP. Finally, we get the conversion factor of $F_{\rm UVW2}=4.45\times10^{-5}$ mJy count$^{-1}$. Subsequently, we perform aperture photometry on the upper and lower blobs with a radius of $\sim$1''.5. Due to the pollution of the light from disk in the aperture, we exclude the pixel with the highest flux. The photometric result is shown as red stars in Fig. \ref{fig:bpt_uv}c and d.

\section{Data analysis and result} \label{sec:result}

\subsection{UV-H$\alpha$ relationship of individual pixels}\label{subsec:uv-halpha_relation}

\begin{figure*}[ht]
\centering
\includegraphics[width=1.0\textwidth]{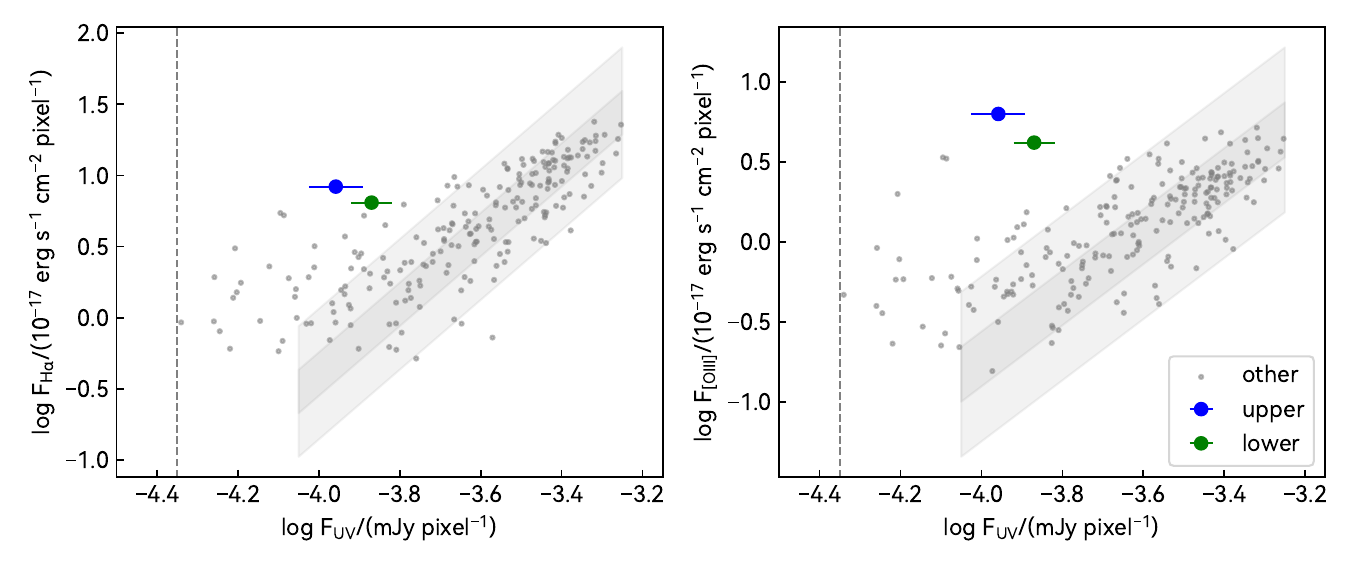}
\caption{Left panel: the flux of H$\alpha$ and UVOT UVW2 band. Right panel: the flux of [O III]$\lambda$5007 and UVOT UVW2 band. The blue dot is average of the resampled pixels of the upper blob. The green dot is the average of the resampled pixels of the lower blob. The gray dots are other pixels. The semi-transparent gray area represents the scatter of relationship fitted by gray dots. The darker area is 1$\sigma_{\rm int}$ range, while the lighter one is 3$\sigma_{\rm int}$ range. The vertical dashed gray line represent the standard deviation of the background of UVOT UVW2 band.}\label{fig:uv_ha_compare}
\end{figure*}

The flux of UV and H$\alpha$ are both the indicator of SFR, and there should be a close relationship between them \citep{Kennicutt1998}. In order to compare the flux of UVOT UVW2 and the emission line maps pixel by pixel, we use the \texttt{reproject} \citep{reproject} code to resample the emission line maps. The pixel-by-pixel comparison between the flux of UVOT UVW2 and the flux of H$\alpha$ and [O III] emission line is shown in Fig. \ref{fig:uv_ha_compare}. We have checked the precision of the astrometry by the stars in the field of view of UVOT, finding that the deviation is less than one pixel (1''). Due to the similar FWHM ($\sim$2''.5) of the point spread function (PSF) of MaNGA and UVOT UVW2 band, we do not further conduct PSF matching.

We use \texttt{emcee} to perform linear fitting on these two relationships, excluding pixels occupied by the two blobs. The residuals between the observed values and the fitted values are contributed by two components: measurement error and the intrinsic scatter ($\sigma_{\rm int}$) of the relationship. We set $\sigma_{\rm int}$ as a free parameter and finally get the $\sigma_{\rm int}$ of the H$\alpha$-UV relationship is $\sim$0.15, while the $\sigma_{\rm int}$ of the [O III]-UV relationship is $\sim$0.17. From this figure, we can see that almost all pixels of the upper and lower blobs are located outside the 3$\sigma_{\rm int}$ of the relationship formed by other pixels in both H$\alpha$-UV and [O III]-UV relationships. This suggests that their physical nature may differ from the star-forming regions on the disk.

\subsection{Gas-phase metallicity} \label{subsec:metallicity}

\begin{figure}[ht]
\centering
\includegraphics[width=1.0\columnwidth]{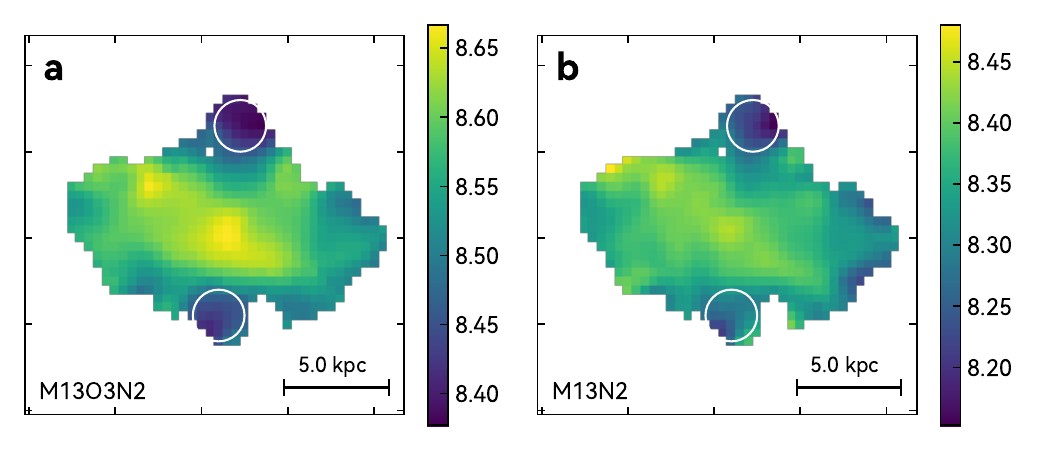}
\caption{The gas-phase metallicity ((12+log(O/H)) map of 9885-9102. a, derived by the O3N2 calibrator \citep{Marino2013}. b, derived by the N2 calibrator \citep{Marino2013}. }\label{fig:metal_map}
\end{figure}

We perform a preliminary gas-phase metallicity measurement, assuming that the whole galaxy is dominated by H II regions. Fig. \ref{fig:metal_map} shows the metallicity (12+log(O/H)) map of this galaxy. We find that the metallicities in these two blobs may be abnormally low. Moreover, it is crucial to point out that the metallicity from the AGN model described in Sect. \ref{subsec:cloudy} is even lower than the values shown in Fig. \ref{fig:metal_map}, suggesting the external origin of the gas of blobs.

\subsection{Spectral extraction and fitting} \label{subsec:fitting}

The spectra of the blobs are extracted in a circular aperture with a radius of 1''.5 from the MaNGA DAP datacube. Red circles shown in Fig. \ref{fig:main_image} are the extracted area. We find that the shape of continua of the blobs is very close to the surrounding extra structure. The only difference is that the surrounding structure has no emission lines, only continua spectra, while blobs have both. We subtract the velocity-corrected continuum of the surrounding structure as the background to obtain the pure emission line spectrum for the two blobs respectively. The aperture of the background is the same as the extraction of the blobs. We adopt the same method as \cite{Yao2022} to fit emission lines, using the \texttt{MPFIT} code. The $n_e$ and the 1$\sigma$ upper limit of T$_e$ are calculated by the \texttt{PyNeb} \citep{pyneb} code.

\subsection{Photoionization models} \label{subsec:cloudy}

We construct detailed photoionization models for both star formation and AGN, to quantitatively examine which model is preferred to match the emission line ratios from the MaNGA survey and the UV continuum from Swift observation. We adopt \texttt{Cloudy} \citep{cloudy} code to fit the four line ratios, [O II]/H$\beta$, [O III]/H$\beta$, [N II]/H$\alpha$, and [S II]/H$\alpha$, from the measurements in Sect. \ref{subsec:fitting}. These four line ratios, including the two used in the BPT diagram, are commonly used to determine the properties of ionized plasma \citep{PerezMontero2014,Kewley2019}. The detailed information of the incident spectra we used in \texttt{Cloudy} is described in Appendix \ref{subapdx:sed_shape}. 
Additionally, due to the weak dust extinction in the two blobs, with an H$\alpha$/H$\beta$ ratio of  $\sim$3 (see Fig. \ref{fig:kinematic}d), and the close wavelengths of the relevant lines for the above four line ratios, the input flux ratios presented are uncorrected for dust extinction. Even for the [O II]/H$\beta$ ratio, the impact of dust is less than 10\% according to the \cite{Calzetti2000} dust-attenuation law.

We use the Markov-chain Monte-Carlo (MCMC) method provided by \texttt{emcee}\citep{emcee} to explore the parameter space of Oxygen abundance, Nitrogen abundance, and ionization parameter. Technically, \texttt{emcee} interacts with \texttt{Cloudy} through the \texttt{pyCloudy} \citep{pycloudy} package. The range of 12+log(O/H) is limited between 7.6 and 8.6. The range of logU is limited between -4.0 and -2.0. The range of log(N/O) is limited between -1.5 to 0.0. For the AGN model of ``Free $\alpha$'', the range of $\alpha_{\rm OX}$ is limited between $-1.8$ and $-0.8$, and the range of $\alpha_{\rm UV}$ is limited between $-2.0$ and $-0.4$. The geometry is set to be plane-parallel, and the calculation stops when the column density reaches 10$^{21}$ cm$^{-2}$. All other metal abundances except the nitrogen abundance are scaled to oxygen following the solar proportions. The helium abundance is set according to \cite{Dopita2006}. We use the line ratio of [S II]$\lambda$6716/[S II]$\lambda$6731 to estimate the electron densities ($n_e$), finding that the $n_e$ are $\sim$15 and $\sim$35 cm$^{-3}$ for the upper and lower blobs. In the run of \texttt{Cloudy}, we adopt a fixed $n_e$ of 20 cm$^{-3}$. The prior probability density of each free parameter is considered to be uniform within the range, while it is 0 outside the range. The likelihood function is obtained from the residuals of four line ratios and their uncertainties.

\begin{deluxetable*}{cccccc}
\tabletypesize{\scriptsize}
\tablewidth{0pt} 
\tablecaption{Summary of the fitting result} \label{tab:summary}
\tablehead{\colhead{SED} & \colhead{12+log(O/H)} & \colhead{logU} & \colhead{log(N/O)} & \colhead{$\alpha_{\rm OX}$} & \colhead{$\alpha_{\rm UV}$}}
\colnumbers
\startdata
5$\times$10$^4$K Blackbody & $8.285_{-0.208}^{+0.116}$ & $-2.627_{-0.068}^{+0.050}$ & $-1.032_{-0.075}^{+0.105}$ & - & - \\
MF87 & $7.829_{-0.027}^{+0.032}$ & $-3.076_{-0.041}^{+0.055}$ & $-0.933_{-0.081}^{+0.111}$ & $-1.400$ & $-0.511$ \\
UVsoft & $7.839_{-0.014}^{+0.015}$ & $-3.015_{-0.015}^{+0.013}$ & $-0.926_{-0.039}^{+0.038}$ & $-1.504$ & $-0.491$ \\
HE0238 & $7.906_{-0.030}^{+0.053}$ & $-2.975_{-0.034}^{+0.039}$ & $-0.991_{-0.068}^{+0.086}$ & $-1.454$ & $-0.500$ \\
Free $\alpha$ & $7.887_{-0.063}^{+0.165}$ & $-2.939_{-0.082}^{+0.126}$ & $-0.961_{-0.115}^{+0.113}$ & $-1.420_{-0.232}^{+0.301}$ & $-1.173_{-0.319}^{+0.311}$ \\
\hline
5$\times$10$^4$K Blackbody & $8.124_{-0.199}^{+0.330}$ & $-2.605_{-0.072}^{+0.098}$ & $-0.858_{-0.134}^{+0.116}$ & - & - \\
MF87 & $7.784_{-0.044}^{+0.053}$ & $-3.040_{-0.060}^{+0.060}$ & $-0.836_{-0.093}^{+0.102}$ & $-1.400$ & $-0.511$ \\
UVsoft & $7.785_{-0.047}^{+0.065}$ & $-2.989_{-0.058}^{+0.059}$ & $-0.862_{-0.107}^{+0.111}$ & $-1.503$ & $-0.489$ \\
HE0238 & $7.843_{-0.051}^{+0.070}$ & $-2.934_{-0.060}^{+0.074}$ & $-0.844_{-0.120}^{+0.103}$ & $-1.454$ & $-0.500$ \\
Free $\alpha$ & $7.828_{-0.067}^{+0.177}$ & $-2.911_{-0.100}^{+0.147}$ & $-0.861_{-0.151}^{+0.134}$ & $-1.399_{-0.282}^{+0.331}$ & $-1.209_{-0.343}^{+0.349}$ \\
\enddata
\tablecomments{The upper group in the table is the fitting result of the upper blob, and the lower group is the fitting result of the lower blob. The value itself, value in the lower right, and value in the upper right represent the median, the 16\% and 84\% quantiles of the posterior distribution, respectively. The $\alpha_{\rm OX}$ and $\alpha_{\rm UV}$ of MF87, UVsoft and HE0238 without quantiles are fixed values.}
\end{deluxetable*}

The median and $\pm$1$\sigma$ quantile of each parameter derived by the \texttt{Cloudy} and \texttt{emcee} are listed in Table \ref{tab:summary}. For the AGN model with two additional free parameters, we find that adding these parameters has little effect on the posterior distributions of log(O/H), logU, and log (N/O). The median of $\alpha_{\rm OX}$ is consistent with the first three fixed-shape AGN spectra, but $\alpha_{\rm UV}$ is almost not constrained by these sets of line ratios.

After obtaining the posterior distribution of each parameter, we take this posterior distribution as input of \texttt{Cloudy} and recalculate the line ratios of five photoionization models with different incident spectra. We find that all these models can well reproduce the observed line ratios (see Fig. \ref{fig:line_ratio} in Appendix \ref{apdx:emcee}). This indicates that the line ratios solely can not separate the two models in case of low metallicity \citep{Reines2013,Kimbro2021,Uebler2023,Zhu2023}.

We generate the grids with different incident spectra on the [N II] and [S II] BPT diagrams at varying 12+log(O/H), logU. The range of 12+log(O/H) is 7.6 to 8.4, with an interval of 0.2. The range of logU is $-4.0$ to $-2.5$ with an interval of 0.5. The log(N/O) is set to a fixed value of $-1.0$ according to the posterior result in Table \ref{tab:summary}. We find that all AGN-type grids (MF87, UVsoft, and HE0238) resemble each other. The grid of MF87 is shown in Fig. \ref{fig:bpt_uv} as an example.

\section{Discussion} \label{sec:discuss}

\subsection{The origin of the bipolar H$\alpha$ blobs}

As Sect. \ref{subsec:swift} mentioned, we find that no X-ray photon is detected (3$\sigma$ limit of L$_{\rm 0.3-10\ keV}$ is 4.3$\times10^{41}$ erg s$^{-1}$). This suggests the center of MaNGA 9885-9102 may not host an AGN. However, the X-ray photons of 0.3-10 Kev band can be easily obscured by gas and dust \citep{Xue2017}, keeping the possibility of an obscured AGN below the detection limit. Overall, we can not give a firm conclusion solely based on the soft X-ray observation whether this galaxy hosts an AGN or not.

\begin{figure*}[ht]
\centering
\includegraphics[width=1\textwidth]{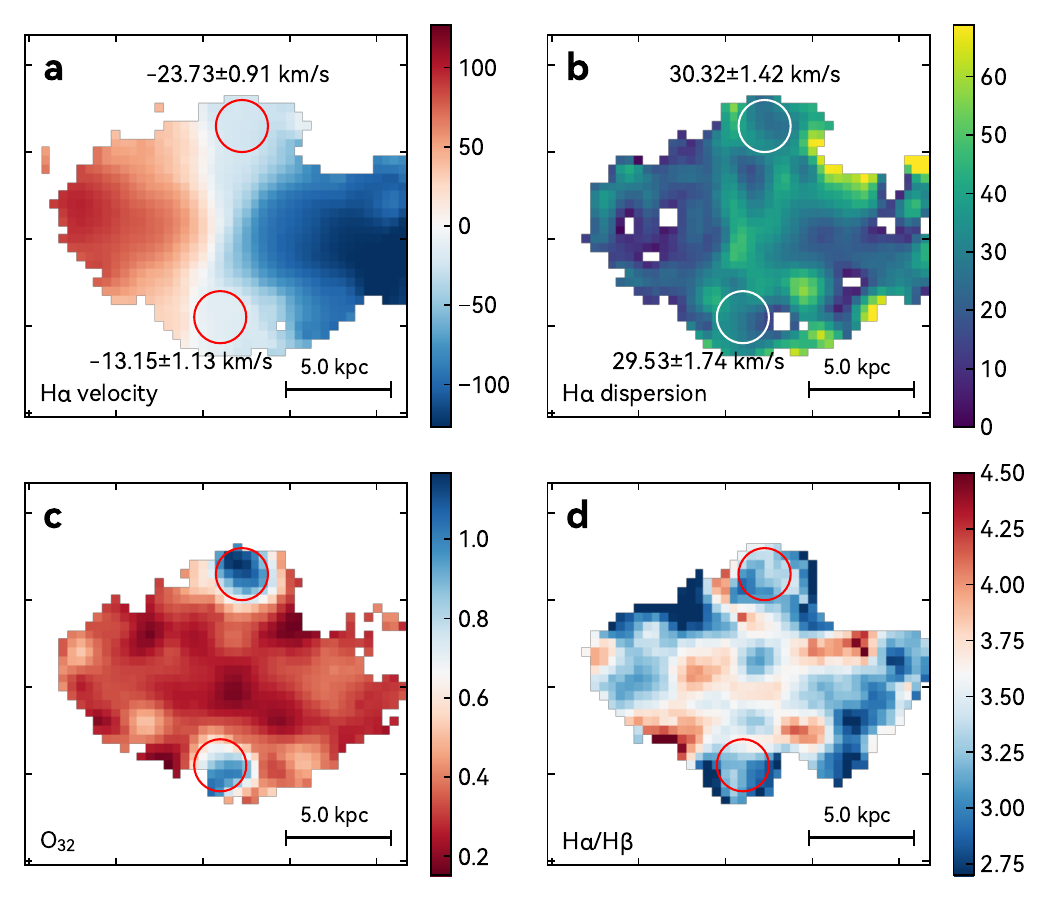}
\caption{The properties maps of MaNGA 9885-9102. Panel a: the velocity map of emission lines. Panel b: the velocity dispersion map of emission lines. All data are taken from the DAP. The value shown beside the blobs is the average and its uncertainty of pixels in the circle. Panel c: the map of [O III]$\lambda$5007/[O II]$\lambda$3727 flux ratios. Panel d: the map of H$\alpha$/H$\beta$ flux ratios. }
\label{fig:kinematic}
\end{figure*}

However, in the UVOT UVW2 image, there is also no UV counterpart for the blobs on with a visual inspection, as shown in Fig. \ref{fig:main_image}f. On the contrary, the other regions with strong H$\alpha$ emission for this galaxy show evident counterparts on the UVOT UVW2 image. We have examined that the absence of UV counterpart for blobs is not due to dust obscuration. The map of H$\alpha$/H$\beta$ ratio of this galaxy, shown in Fig. \ref{fig:kinematic}d, indicates that the dust extinction of the blobs is very weak under Case-B recombination \citep{Hummer1987}.

The absence of UV counterparts of the blobs indicates that the ionizing source may not be inside the blob, which rejects the star-forming model. Actually, the UV and H$\alpha$ luminosity both represent the star formation activities, and a close positive correlation between the two is anticipated for star-forming regions in Fig. \ref{fig:uv_ha_compare} \citep{Kennicutt1998,salim_sfms}. If the source of ionization is external, we can only observe emission lines, weak recombination and diffusion continuum from the nebula, excluding the ionizing field from the source. Combining this result with the symmetric geometry of the two blobs at the minor axis, we qualitatively draw the conclusion that the blobs are probably from AGN photoionization.

The Oxygen abundance, $12 + \log({\rm O/H})$, constraint from the AGN model is $\sim$7.8 for the two blobs. This low metallicity, with respect to H II regions of this galaxy indicated by Fig. \ref{fig:kinematic}c, suggests that the ionized gas may not originally come from the central region of this galaxy, since the outflowing gas driven by AGN is usually metal-rich.  In consistent with this, these two blobs do not show significant velocity offset ($\sim$20 km s$^{-1}$) seen from Fig. \ref{fig:kinematic}a, indicating they are not moving fast away from the galaxy. Fig. \ref{fig:kinematic}b shows that the velocity dispersion of H$\alpha$ is not high, which can rule out the ionization from shocks \citep{Rich2011,Rich2015,Ji2021}. Because shocks can produce a much higher velocity dispersion of H$\alpha$, with 150-500 km s$^{-1}$ \citep{Kewley2019}.

\subsection{Luminosity estimation} \label{subsec:lum_estimate}

Here we estimate the luminosity through the photoionization models in the both cases of AGN and star formation. The definition of ionization parameter (U) is
\begin{equation}
U\equiv\frac{Q(H)}{4\pi{}R^2n_{\rm H}c}, \label{eq:U}
\end{equation}
where Q(H) is the total number of ionizing photons per unit time emitted by the source, R is the distance from the nebula to the ionizing source and c is the speed of light. The Q(H) can be derived by
\begin{equation}
Q(H)=\int_{\nu_1}^{\nu_2}\frac{L_\nu}{h\nu}d\nu, \label{eq:Q}
\end{equation}
where L$_\nu$ is the absolute luminosity per unit frequency of the source. The default energy range (h$\nu_1$ to h$\nu_2$) in \texttt{Cloudy} is 1 Ryd to 7.354$\times$10$^6$ Ryd.

For any brightness of source, as long as the spectral shape is the same, the ratio of Q(H) to the total bolometric luminosity (L$_{bol}$) is also the same. So, the L$_{bol}$ is
\begin{equation}
L_{\rm bol}=\int_{0}^{+\infty}L_\nu d\nu=\frac{\int_{0}^{+\infty}F_\nu d\nu}{\int_{\nu_1}^{\nu_2}\frac{F_\nu}{h\nu}d\nu}Q(H), \label{eq:L}
\end{equation}
where F$_\nu$ is the flux per unit frequency of the shape of spectrum of the source. The coefficient of Q(H) (i.e. the ratio of two integrals) represents the total energy required to produce each ionizing photon. For the spectral shape of MF87, UVsoft, HE0238, and blackbody (5$\times$10$^4$K), this energy is 69.6, 76.9, 103.2, and 35.3 eV, respectively.

\subsubsection{Black hole mass and Eddington ratio for the AGN model} \label{subsubsec:edd_estimate}

The mass of central black hole ($M_{\rm BH}$) for MaNGA 9885-9102 is estimated from the central velocity dispersion ($\sigma_*$) following the relation \citep{coevolution_araa}:
\begin{equation}
\log{(M_{\rm BH}/M_\odot)}=8.46 + 4.26 \log(\sigma_*/{\rm 200\ km\ s^{-1}}). \label{eq:MBH_sigma}
\end{equation}

The $\sigma_*$ corrected by the instrumental broadening is $\sim$49$\pm$9 km s$^{-1}$. We obtain $M_{\rm BH}\sim$7.2$\times 10^5M_\odot$ with an intrinsic uncertainty of 0.3 dex from the relation. The Eddington luminosity limit is $\log{[L_{\rm Edd}/(\rm erg~s^{-1})]}\simeq43.96\pm0.3$. This means that MaNGA 9885-9102 likely hosts an IMBH at the center, which is responsible for the two blobs discussed above.

However, we analyze the surface brightness profile measured in Appendix \ref{apdx:desi_profile} can be well characterized by a single-exponential function, without evidence of significant bulge, which means that using this relationship to estimate the $M_{\rm BH}$ can be inappropriate, but this does not prevent the possibility of the existence of an IMBH in MaNGA 9885-9102.

Another approach to estimate its M$_{\rm BH}$ is through its $M_*$ \citep{mbh_mass}:
\begin{equation}
\log{(M_{\rm BH}/M_\odot)}=(7.45\pm0.08) + (1.05\pm0.11)\log{(M_*/10^{11}M_\odot)}, \label{eq:MBH_mass}
\end{equation}
with an intrinsic scatter of 0.24 dex. Thus, we can get $\log{(M_{\rm BH}/M_\odot)}\simeq$6.02$\pm$0.28 and $\log{[L_{\rm Edd}/(\rm erg~s^{-1})]}\simeq44.12\pm0.28$.

The projection-corrected distance from the center of this galaxy to the blob is $\sim$4.5kpc. Since the $\log U$ given by the three AGN spectra are both around $-3$, we directly substitute it into Eq. \ref{eq:U}. The total energy required to produce each ionizing photon in Eq. \ref{eq:L} we adopt is the average of the three AGN spectra of MF87, UVsoft, and HE0238. Thus, we can obtain $\log{[L_{\rm bol}/(\rm erg~s^{-1})]}\simeq$44.26, assuming isotropic radiation of AGN. This value is close to L$_{\rm Edd}$ of the black hole, indicating $\lambda_{\rm Edd}\gtrsim$70\%. This black hole may be highly active now or in the past. However, we need to emphasize that the estimation of mass and luminosity of the black hole mentioned above is highly uncertain, so the obtained $\lambda_{\rm Edd}$ is only a rough lower limit. 

We use the same method to estimate the luminosity and SFR under the case of central starburst using the blackbody SED rather than the AGN, finding that the SFR at the central region is $\gtrsim 1 M_\odot$ yr$^{-1}$. This SFR is too high to be compatible with overall SFR shown in Fig. \ref{fig:sfms} and Tab. \ref{tab:info}. Therefore, we can conclude that the ionizing source at its center is more likely to be an AGN.

\subsubsection{UV flux estimation} \label{subsubsec:uv_estimate}

We estimate the flux observed in the UVOT UVW2 band for the star formation (in Fig.\ref{fig:shiyitu}a) and the AGN model (in Fig.\ref{fig:shiyitu}b) respectively. We consider the emission nebula as a sphere with a radius of 0.5 kpc. And use the posterior probability density of each parameter to calculate the distribution of total luminosity of the blobs in the UVW2 band through \texttt{Cloudy}. We use the \cite{Calzetti2000} dust-attenuation law and compare the observed Balmer decrement (H$\alpha$/H$\beta$ ratio) to the intrinsic values of 2.86 \citep{Hummer1987} under the case-B assumption to correct the dust extinction of UV flux. We use the Monte Carlo method to calculate the distribution of UV flux in the UVW2 band. We create mock image of the two blobs using the dust-corrected UV flux estimated by the star formation model in the Fig. \ref{fig:uv_mock}, which shows that we should see obvious features of the blobs on the UV image in case of star formation origin, but we cannot actually. 

Furthermore, in the lower two panels of Fig. \ref{fig:bpt_uv}, the mean value is shown as colored horizontal line and the standard deviation (i.e. the uncertainty) is shown as colored area. For the star-forming model, the UV luminosity is mainly contributed by the ionizing source (a blackbody with T=5$\times$10$^4$K) itself. For the AGN model, the UV luminosity is mainly contributed by the diffuse continuum. We calculate the diffuse continuum for the AGN incident spectrum of MF87, UVsoft, and HE0238 respectively. There is no significant difference in their results. The green areas and lines shown in the lower two panels of Fig. \ref{fig:bpt_uv} are the average result of three AGN models. 
We notice that the flux in the lower blob is slightly higher than the AGN model, though this discrepancy is within 1$\sigma$. This  might result from the low resolution of the UVW2 image, potentially causing UV emission in aperture photometry to be affected by the galaxy disk. Another possibility is a hybrid ionization mechanism involving multiple sources. If both star formation and the AGN contribute to the UV flux, about 10\% of the lower blob's UV flux could originate from star formation, though this contribution is relatively weak. Therefore, if ionizing photons partially come from star formation, the Eddington ratio could be slightly lower.

The significant deviation ($>$4$\sigma$) between the observed UV flux and the predicted values of the star-forming model poses a challenge for using in-situ star formation to explain the physical origin of these two blobs due to their low UV luminosity, while the AGN models successfully reproduce the observed flux of UVOT UVW2 within 1$\sigma$, verifying the AGN origin.

\section{Conclusion} \label{sec:conclusion}

The case study in this work confirms that such an active AGN from probable IMBH is neglected by the traditional BPT diagram. The off-centered H$\alpha$ blobs within galaxies are usually treated as H II regions by the BPT diagram without considering different possibilities. New methods are required to identify the light echoes of AGN, especially for low-mass galaxies. In this work, we present a new method, i.e. utilizing the UV continuum to specify whether the source of ionization is within or out of the off-centered H$\alpha$ blobs. The bipolar signature is not necessary for this method. It is critical for this method to search for the evidences of IMBHs because IMBHs usually reside in low-mass and even dwarf galaxies with very low metallicity.  We plan to apply this method to the GALEX survey in the future to explore the observational evidences of low-to-intermediate black holes.  This approach likely opens a new window to explore the radiative feedback of AGN for low-to-intermediate black holes.

\begin{acknowledgments}
Funding for the Sloan Digital Sky Survey IV has been provided by the Alfred P. Sloan Foundation, the U.S. Department of Energy Office of Science, and the Participating Institutions. SDSS acknowledges support and resources from the Center for High-Performance Computing at the University of Utah. The SDSS web site is \url{www.sdss4.org}.

SDSS is managed by the Astrophysical Research Consortium for the Participating Institutions of the SDSS Collaboration including the Brazilian Participation Group, the Carnegie Institution for Science, Carnegie Mellon University, Center for Astrophysics | Harvard \& Smithsonian (CfA), the Chilean Participation Group, the French Participation Group, Instituto de Astrofísica de Canarias, The Johns Hopkins University, Kavli Institute for the Physics and Mathematics of the Universe (IPMU) / University of Tokyo, the Korean Participation Group, Lawrence Berkeley National Laboratory, Leibniz Institut für Astrophysik Potsdam (AIP), Max-Planck-Institut für Astronomie (MPIA Heidelberg), Max-Planck-Institut für Astrophysik (MPA Garching), Max-Planck-Institut für Extraterrestrische Physik (MPE), National Astronomical Observatories of China, New Mexico State University, New York University, University of Notre Dame, Observatório Nacional / MCTI, The Ohio State University, Pennsylvania State University, Shanghai Astronomical Observatory, United Kingdom Participation Group, Universidad Nacional Autónoma de México, University of Arizona, University of Colorado Boulder, University of Oxford, University of Portsmouth, University of Utah, University of Virginia, University of Washington, University of Wisconsin, Vanderbilt University, and Yale University.

The Legacy Surveys consist of three individual and complementary projects: the Dark Energy Camera Legacy Survey (DECaLS; Proposal ID \#2014B-0404; PIs: David Schlegel and Arjun Dey), the Beijing-Arizona Sky Survey (BASS; NOAO Prop. ID \#2015A-0801; PIs: Zhou Xu and Xiaohui Fan), and the Mayall z-band Legacy Survey (MzLS; Prop. ID \#2016A-0453; PI: Arjun Dey). DECaLS, BASS and MzLS together include data obtained, respectively, at the Blanco telescope, Cerro Tololo Inter-American Observatory, NSF’s NOIRLab; the Bok telescope, Steward Observatory, University of Arizona; and the Mayall telescope, Kitt Peak National Observatory, NOIRLab. Pipeline processing and analyses of the data were supported by NOIRLab and the Lawrence Berkeley National Laboratory (LBNL). The Legacy Surveys project is honored to be permitted to conduct astronomical research on Iolkam Du’ag (Kitt Peak), a mountain with particular significance to the Tohono O’odham Nation.

NOIRLab is operated by the Association of Universities for Research in Astronomy (AURA) under a cooperative agreement with the National Science Foundation. LBNL is managed by the Regents of the University of California under contract to the U.S. Department of Energy.

This project used data obtained with the Dark Energy Camera (DECam), which was constructed by the Dark Energy Survey (DES) collaboration. Funding for the DES Projects has been provided by the U.S. Department of Energy, the U.S. National Science Foundation, the Ministry of Science and Education of Spain, the Science and Technology Facilities Council of the United Kingdom, the Higher Education Funding Council for England, the National Center for Supercomputing Applications at the University of Illinois at Urbana-Champaign, the Kavli Institute of Cosmological Physics at the University of Chicago, Center for Cosmology and Astro-Particle Physics at the Ohio State University, the Mitchell Institute for Fundamental Physics and Astronomy at Texas A\&M University, Financiadora de Estudos e Projetos, Fundacao Carlos Chagas Filho de Amparo, Financiadora de Estudos e Projetos, Fundacao Carlos Chagas Filho de Amparo a Pesquisa do Estado do Rio de Janeiro, Conselho Nacional de Desenvolvimento Cientifico e Tecnologico and the Ministerio da Ciencia, Tecnologia e Inovacao, the Deutsche Forschungsgemeinschaft and the Collaborating Institutions in the Dark Energy Survey. The Collaborating Institutions are Argonne National Laboratory, the University of California at Santa Cruz, the University of Cambridge, Centro de Investigaciones Energeticas, Medioambientales y Tecnologicas-Madrid, the University of Chicago, University College London, the DES-Brazil Consortium, the University of Edinburgh, the Eidgenossische Technische Hochschule (ETH) Zurich, Fermi National Accelerator Laboratory, the University of Illinois at Urbana-Champaign, the Institut de Ciencies de l’Espai (IEEC/CSIC), the Institut de Fisica d’Altes Energies, Lawrence Berkeley National Laboratory, the Ludwig Maximilians Universitat Munchen and the associated Excellence Cluster Universe, the University of Michigan, NSF’s NOIRLab, the University of Nottingham, the Ohio State University, the University of Pennsylvania, the University of Portsmouth, SLAC National Accelerator Laboratory, Stanford University, the University of Sussex, and Texas A\&M University.

BASS is a key project of the Telescope Access Program (TAP), which has been funded by the National Astronomical Observatories of China, the Chinese Academy of Sciences (the Strategic Priority Research Program “The Emergence of Cosmological Structures” Grant \# XDB09000000), and the Special Fund for Astronomy from the Ministry of Finance. The BASS is also supported by the External Cooperation Program of Chinese Academy of Sciences (Grant \# 114A11KYSB20160057), and Chinese National Natural Science Foundation (Grant \# 12120101003, \# 11433005).

The Legacy Survey team makes use of data products from the Near-Earth Object Wide-field Infrared Survey Explorer (NEOWISE), which is a project of the Jet Propulsion Laboratory/California Institute of Technology. NEOWISE is funded by the National Aeronautics and Space Administration.

The Legacy Surveys imaging of the DESI footprint is supported by the Director, Office of Science, Office of High Energy Physics of the U.S. Department of Energy under Contract No. DE-AC02-05CH1123, by the National Energy Research Scientific Computing Center, a DOE Office of Science User Facility under the same contract; and by the U.S. National Science Foundation, Division of Astronomical Sciences under Contract No. AST-0950945 to NOAO.

We thank the Swift Science Operations team for accepting our ToO requests and arranging the observations.
\end{acknowledgments}

%

\vspace{5mm}
\facilities{DESI, Sloan(MaNGA), Swift(XRT and UVOT)}


\software{
    astropy \citep{Collaboration2022},
    photutils \citep{photutils},
    reproject \citep{reproject},
    Cloudy \citep{cloudy,pycloudy},
    statmorph\_csst \citep{statmorph,statmorph_csst},
    mpfit \citep{mpfit},
    emcee \citep{emcee},
    PyNeb \citep{pyneb}
}



\appendix

\section{Scaling relationships} \label{apdx:sfms}

\begin{figure}[ht]
\centering
\includegraphics[width=1.0\textwidth]{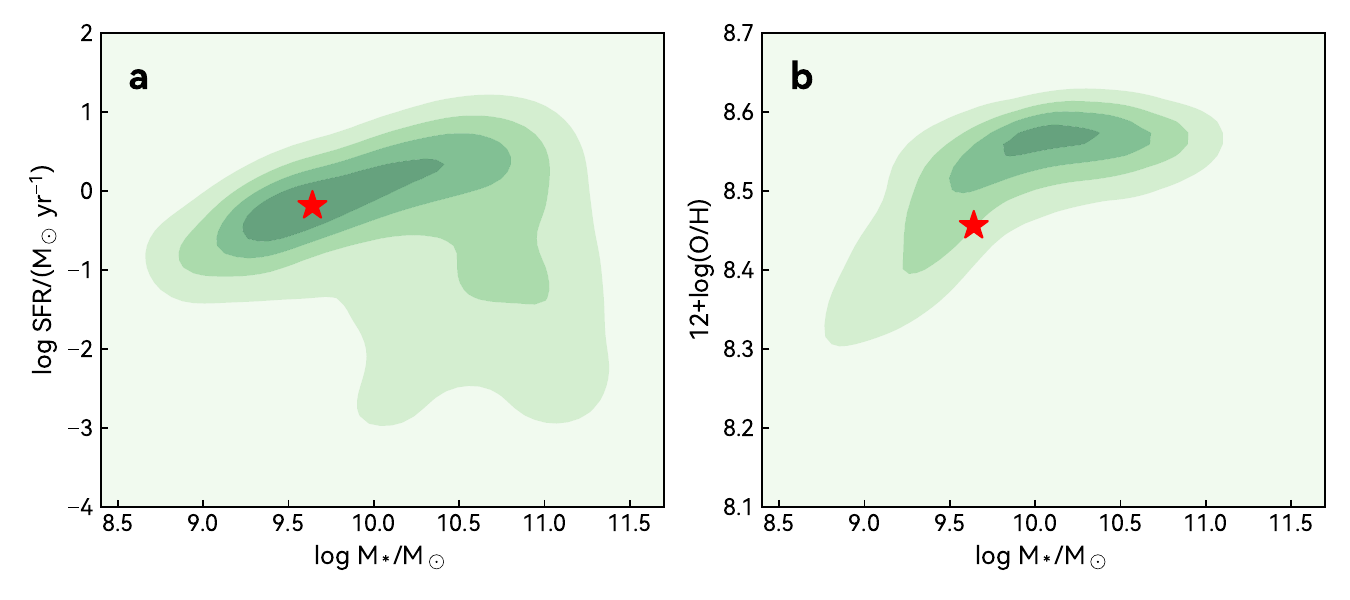}
\caption{The scaling relationships of 9885-9102. a, the star formation main sequence. b, the mass-metallicity relation. The red stars represent 9885-9102, and the green areas represent the distribution of other MaNGA galaxies.}\label{fig:sfms}
\end{figure}

We examine the basic scaling relations of SFR-$M_*$ and metallicity-$M_*$ for this galaxy, shown in Fig. \ref{fig:sfms}.  The SFR measurements are based on the H$\alpha$ and 12+log(O/H) uses the {\tt O3N2} calibrator at the central region. The reference sample is taken from the same catalog \citep{pipe3d} with available H$\alpha$-based SFR. As shown, the MaNGA 9885-9102 appears to be a normal star-forming galaxy on SFR-$M_*$ diagram with slightly low Oxygen abundance with respect to the counterparts of the same stellar mass.  

\section{The surface brightness profile measurement} \label{apdx:desi_profile}

\begin{figure}[ht]
\centering
\includegraphics[width=0.75\columnwidth]{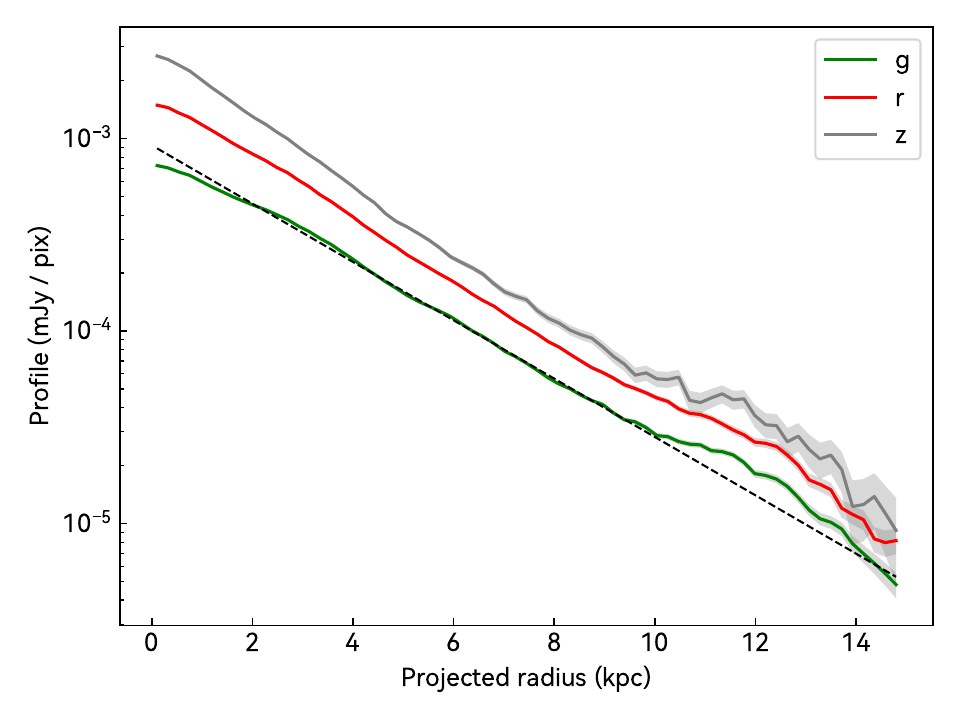}
\caption{Surface brightness profiles in the DESI $g$,$r$,$z$ bands of MaNGA 9885-9102. The black dashed black line is the exponential fitting of the g-band image. The horizontal axis is the projection corrected distance to center. }\label{fig:desi_profile}
\end{figure}

The surface brightness profiles of Fig. \ref{fig:desi_profile} are measured using \texttt{photutils} \citep{photutils} in the DESI Legacy Survey $g$,$r$,$z$ bands. The unrelated surrounding sources are masked manually. The photometry adopts elliptical apertures, whose azimuth and $b/a$ come from the DESI Tractor catalog. The length of the DESI image pixel is 0''.262, corresponding projected physical length of $\sim$0.216 kpc.

Interestingly, we find that the surface brightness profiles in all bands have exponential form, while an excess of light can be seen at the radius of $\sim$12 kpc. This may be related to gas accretion by capturing dwarf galaxies, which may also be responsible for the origin of the gas of blobs. Deeper observations are needed to approve and disprove this hypothesis.

\section{Additional information about the photoionization models}\label{apdx:emcee}

\subsection{The shape of the incident spectra} \label{subapdx:sed_shape}

\begin{figure}[ht]
\centering
\includegraphics[width=0.75\columnwidth]{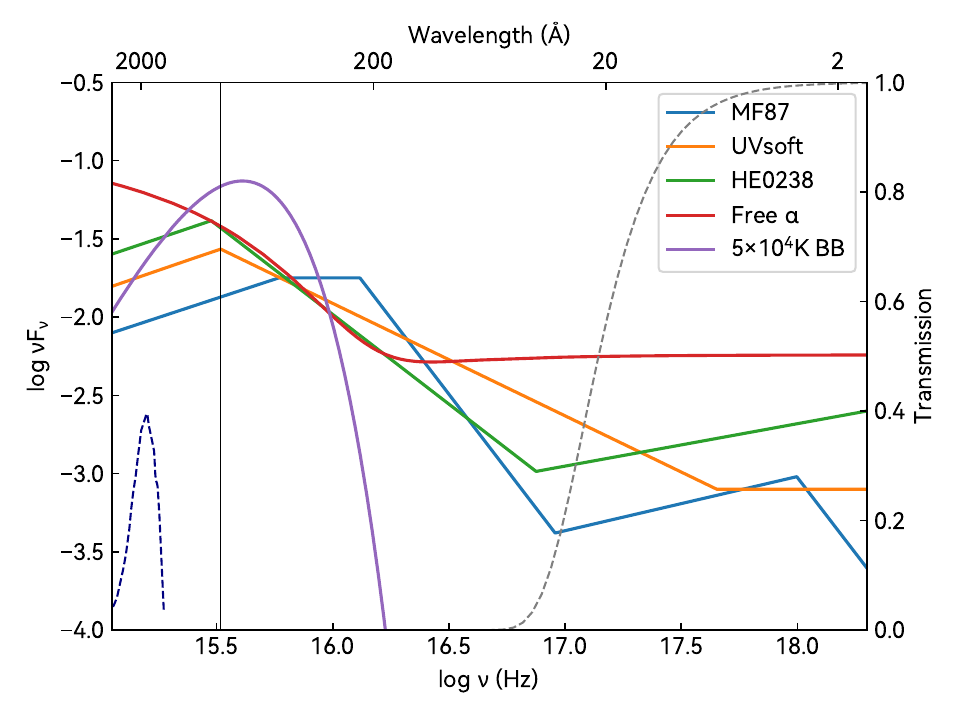}
\caption{The incident spectra of MF87, UVsoft, HE0238, Free $\alpha$, and 5$\times$10$^4$K blackbody. The $\alpha_{\rm OX}$ and $\alpha_{\rm UV}$ of the ``Free $\alpha$'' spectra adopt the values of the upper blob. The vertical solid black line represents the first ionization energy of hydrogen (i.e. 13.6 eV or 1 Ryd), the dashed dark blue curve represents the transmission of UVW2 band of UVOT, and the dashed gray curve represents the transmission of neutral hydrogen with N$_{H}$=10$^{21}$ cm$^{-2}$.} \label{fig:incident}
\end{figure}

For the AGN model, we adopt four types of incident spectrum as the ionization source. The first three types are MF87 \citep{mf87}, UVsoft \citep{uvsoft}, and HE0238 \citep{he0238}, respectively. The last one ``Free $\alpha$'' has two additional free parameters, $\alpha_{\rm OX}$ and $\alpha_{\rm UV}$, using the \texttt{Cloudy} command of ``AGN T=2e5 k, a(ox)=aox, a(uv)=auv''. The $\alpha_{\rm OX}$ is the spectral index describing the continuum between X-ray (2keV) and UV (2500\AA) \citep{Tananbaum1979}, and the $\alpha_{\rm UV}$ is the spectral index of the Big Bump continuum. For the star formation model, we adopt the blackbody (5$\times$10$^4$K) represents the spectrum of the young stellar population. The shape of each incident spectrum is shown in Fig. \ref{fig:incident}.

\subsection{The posterior line ratios} \label{subapdx:line_ratio}

\begin{figure*}[ht]
\centering
\includegraphics[width=1.0\textwidth]{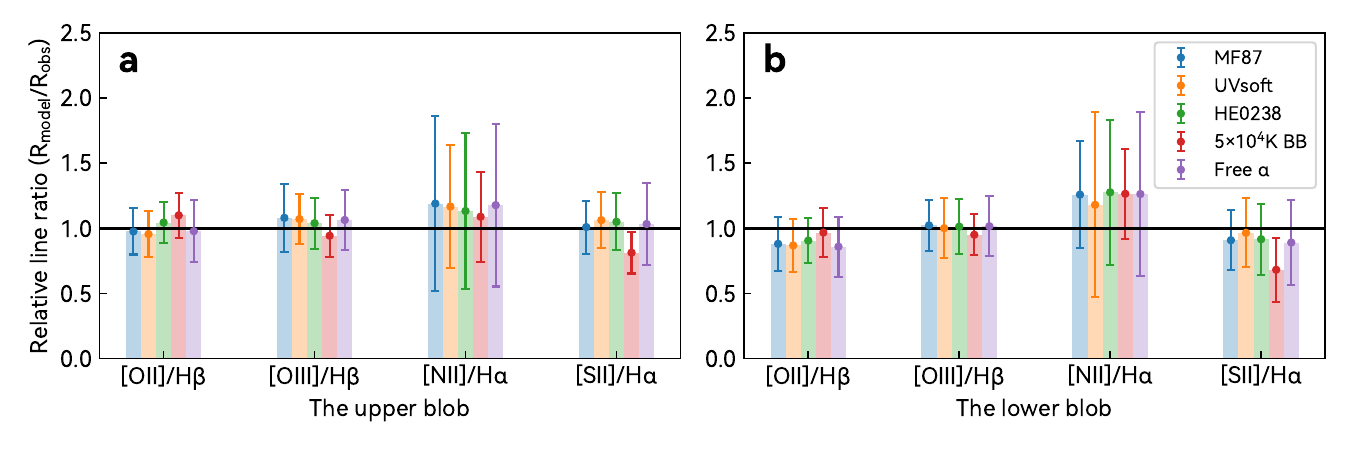}
\caption{The relative values of the line ratios from photoionization models (R$_{\rm model}$) to the observed line ratios (R$_{\rm obs}$). The closer it is to 1 (the horizontal solid black line), the more the model conforms to the observation. Different color of bars represents the different incident spectrum used in the photoionization model.}\label{fig:line_ratio}
\end{figure*}

Fig. \ref{fig:line_ratio} shows the comparison between the line ratios of the models and the observed ratios. The error bars in the figure include the standard deviation of the line ratio of photoionization models generated by the posterior distribution and the uncertainty of the observed values.

\subsection{The UV visibility in the case of star formation} \label{subapdx:uv_mock}

\begin{figure*}[ht]
\centering
\includegraphics[width=1.0\textwidth]{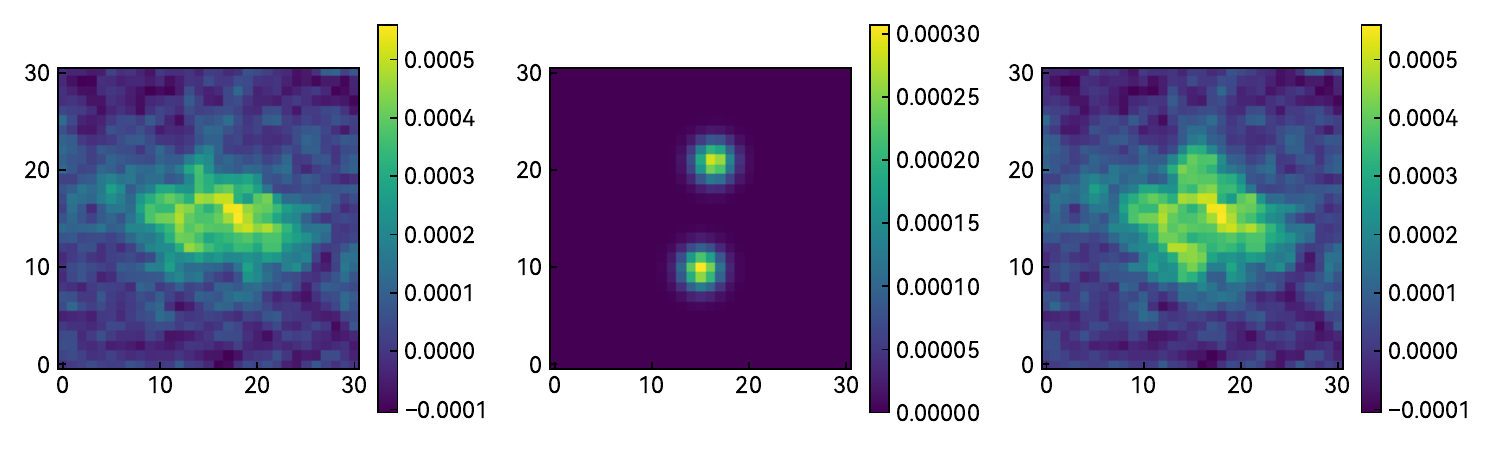}
\caption{Left: The UVOT UVW2 image of 9885-9102. Middle: the mock image of the two blobs with the dust-corrected luminosity estimated by the star formation model. The size is determined by the PSF of UVW2. Right: The image superimposed by the first two. This shows that the obvious features of the blobs on the UV image should be clearly seen in case of star formation origin, but we cannot actually see.}\label{fig:uv_mock}
\end{figure*}

We create mock image of the two blobs using the dust-corrected UV flux estimated by the star formation model in the Fig. \ref{fig:uv_mock} to test the visibility of the two blobs if there are in-situ star formation. Due to the high SFR and weak dust attenuation, we should be able to easily see these two blobs in the UVW2 image (the right panel in Fig. \ref{fig:uv_mock}), even with low spatial resolution and sensitivity.


\bibliography{sample631}{}
\bibliographystyle{aasjournal}



\end{document}